\documentclass[a4paper,11pt]{article}
\usepackage{pos}
\usepackage{placeins}

\title{Normalizing-flow-based density of states for (1+1)D U(1) lattice gauge theory with a $\theta$-term}
\ShortTitle{NF-based DoS for (1+1)D U(1) gauge theory with a $\theta$-term}

\author*[a,b]{Simran Singh}

\author[a,b]{Lena Funcke}

\affiliation[a]{Transdisciplinary Research Area (TRA) Matter, University of Bonn, Germany}
\affiliation[b]{Helmholtz Institute for Radiation and Nuclear Physics (HISKP), University of Bonn, Germany}

\emailAdd{ssingh@uni-bonn.de}
\emailAdd{lfuncke@uni-bonn.de}

\abstract{
A normalizing-flow-based implementation of the density-of-states approach has recently been used to successfully reconstruct the partition function of (1+1)D scalar lattice field theory. In this preliminary work, we extend this framework to a lattice gauge theory by employing gauge-equivariant normalizing flows to reconstruct the density of states of pure (1+1)D U(1) lattice gauge theory, both with and without a $\theta$-term. In the absence of a $\theta$-term, we first demonstrate that the normalizing-flow-based reconstruction of the density of states reproduces the known analytic results for this theory. We further show that, in the presence of a $\theta$-term, this formulation enables the generation of gauge-field configurations at fixed values of the topological charge.
}

\FullConference{The 42nd International Symposium on Lattice Field Theory (LATTICE2025)\\
2-8 November 2025\\
Tata Institute of Fundamental Research, Mumbai, India\\}

\begin{document}
\maketitle

\section{Introduction}

Hybrid Monte Carlo (HMC) simulations sample gauge-field configurations according to 
the path-integral distribution and provide the standard non-perturbative approach for lattice QCD. However, HMC becomes inefficient in several important regimes: simulations near the continuum limit suffer from critical slowing down~\cite{Schaefer:2010hu}, while theories with complex actions, such as those including a topological $\theta$-term, exhibit a numerical sign problem.

The density-of-states (DoS) method offers a promising direction to mitigate both critical slowing down and the sign problem. Recently, normalizing flows (NFs) have been used to compute the DoS directly~\cite{Pawlowski:2022rdn}, rather than reconstructing it via the conventional procedure of measuring and integrating the derivative of its logarithm. In this preliminary work, we extend this approach by employing gauge-equivariant NFs, based on Refs.~\cite{Kanwar:2020xzo,Albergo:2021vyo}, to reconstruct the DoS of pure (1+1)D U(1) lattice gauge theory with and without a $\theta$-term. We first validate the method by comparing the flow-based DoS in the absence of the $\theta$-term with exact results, and then study the theory in the presence of the $\theta$-term, where a sign problem arises.

These proceedings are organized as follows. In Sec.~\ref{sec:1}, we briefly review the DoS method. Section~\ref{sec:2} describes the flow-based DoS approach. In Sec.~\ref{sec:3}, we motivate the choice of (1+1)D U(1) gauge theory as a test case. Results are presented in Sec.~\ref{sec:4}, followed by conclusions in Sec.~\ref{sec:5}.

\section{The density of states} \label{sec:1}

The DoS method was originally introduced to address difficulties encountered in MC simulations of systems undergoing first-order phase transitions \cite{Berg:1991cf,Wang:2000fzi}. In such systems, the probability distribution of the action becomes bimodal, with two peaks corresponding to coexisting phases separated by a region of suppressed probability. As a result, MC configurations tend to become trapped in one peak, leading to inefficient sampling and a loss of ergodicity.

These transitions are associated with changes in the Boltzmann weight controlled by the coupling $\beta$. The DoS method addresses this problem by factorizing the path integral: the $\beta$-independent DoS $\rho(c)$ is computed for fixed slices $c$ of the $\beta$-independent action $S(\phi)$, and the $\beta$-dependence is later reintroduced through a one-dimensional integral over $c$:
\begin{equation}\label{eqn:DoS1}
    \rho(c) = \int \mathcal{D}\phi \, \delta{[c-S(\phi)]} , \qquad Z(\beta) = \int^{c_{\mathrm{max}}}_{c_{\mathrm{min}}} dc \, e^{-\beta c} \rho(c).
\end{equation}
Here, $\phi$ denotes a field configuration (e.g., scalar or gauge fields), $\rho(c)$ counts the number of configurations with constant action $S(\phi)=c$, and $Z(\beta)$ is the canonical partition function.
For most systems of interest, including those considered in this work, the integral over $c$ is bounded within the interval $[c_{\mathrm{min}}, c_{\mathrm{max}}]$.

In this formalism, an observable $O$ is first averaged over all configurations with a fixed action $S(\phi)=c$, giving the slice-dependent expectation value $\langle O \rangle_c$, and then averaged over $c$ to yield the canonical expectation value $\langle O \rangle_\beta$ at coupling $\beta$:
\begin{equation}\label{eqn:DoSOb1}
    \langle O \rangle_c = \frac{1}{\rho(c)}\int \mathcal{D}\phi \, O(\phi) \delta{[c-S(\phi)]} , \qquad    \langle O \rangle_{\beta} = \frac{\int dc \, \rho(c) e^{-\beta c}  \langle O \rangle_c }{\int dc\,  \rho(c) \, e^{-\beta c}}.
\end{equation}
This approach offers two main advantages. First, it eliminates the need to perform separate MC simulations for each value of $\beta$. Second, by factorizing the Boltzmann weight, the coupling $\beta$, which causes the double-peak structure in systems with first-order phase transitions, is decoupled from the sampling procedure.  
This reformulation, however, comes with a trade-off: it requires highly accurate estimates of the density of states $\rho(c)$, particularly near the boundaries of the action slices. Insufficient accuracy in these regions can induce significant errors in the one-dimensional integrals of Eqs.~(\ref{eqn:DoS1}) and (\ref{eqn:DoSOb1}), with the magnitude of the error depending on $\beta$ \cite{Langfeld:2016kty}. 
An extension of the original method \cite{Wang:2000fzi} to quantum field theories with continuous degrees of freedom was proposed in Ref.~\cite{Langfeld:2012ah} and has since been widely applied \cite{Gattringer:2015eey,Gattringer:2016kco,Gattringer:2020mbf,Bennett:2024bhy,Cossu:2021bgn,Lucini:2023irm}. 

A related approach, known as the generalized density of states (gDoS), was introduced in Refs.~\cite{Gocksch:1988iz,Langfeld:2014nta} to tackle the numerical sign problem. Similar to the standard DoS method, the path integral is factorized into two parts. Consider an action with a complex contribution that can be written as $S(\phi) = \beta S_R(\phi) + i h S_I(\phi)$, where $\beta,h\in\mathbb{R}$, and $S_R(\phi)$ and $S_I(\phi)$ denote the real and imaginary parts of the action, respectively. In this case, the gDoS is defined by
\begin{equation}\label{eqn:gDoS1}
    \rho(\beta,c) = \int \mathcal{D}\phi \, e^{-\beta S_R(\phi)} \delta{[c-S_I(\phi)]} \,,\qquad     Z(\beta,h) = \int^{c_{\mathrm{max}}}_{c_{\mathrm{min}}} dc \, e^{-i hc} \rho(\beta,c) \,.
\end{equation}
In this formulation, sampling is performed with respect to the positive semi-definite measure 
$e^{-\beta S_R(\phi)} \, \delta[c - S_I(\phi)]$, while the complex phase enters through the one-dimensional Fourier transform of the gDoS. It is important to emphasize that this procedure \textit{reformulates} rather than solves the sign problem: regions of phase space with strong phase oscillations require highly accurate determinations of $\rho(\beta,c)$ defined in Eq.~\eqref{eqn:gDoS1}. 
Nevertheless, the method has been successfully applied to a variety of theories affected by the sign problem \cite{Gattringer:2015eey,Gattringer:2016kco,Gattringer:2020mbf}, including (1+1)D U(1) gauge theory with a $\theta$-term~\cite{Gattringer:2015eey,Gattringer:2020mbf}, and has even been explored in lattice QCD studies at finite density~\cite{Fodor:2007vv}.

\section{The normalizing-flow-based density of states}\label{sec:2}
NFs have recently been explored for generating gauge configurations in lattice QCD~\cite{Cranmer:2023xbe}, with the goal of overcoming critical slowing down in conventional MC simulations. An NF is an invertible map $f_w$ parametrized by neural networks, which transforms samples $\xi$ drawn from a simple prior distribution $r(\xi)$ into samples $\phi= f_w(\xi)$ of a model distribution $q_w(\phi)$,
\begin{equation}\label{eqn:jacobian}
    q_w(\phi) = r(\xi) \left|\det \left(\frac{\partial f_w}{\partial \xi}\right)\right|^{-1}.
\end{equation}
The parameters $w$ of $f_w$ are optimized so that $q_w(\phi)$ approximates the target distribution, $p(\phi)=e^{-S(\phi)}/Z$, where $S(\phi)$ denotes the action and $Z$ the partition function. As noted in Ref.~\cite{Nicoli:2020njz}, this framework provides a direct estimator for the partition function,
\begin{align}\label{eqn:normDiffpartition}
    Z &= \int \mathcal{D}\phi\, e^{-S(\phi)}
      = \int \mathcal{D}\phi\, q_w(\phi)\frac{e^{-S(\phi)}}{q_w(\phi)}
      = \left\langle e^{-S(\phi)-\log q_w(\phi)} \right\rangle_{\phi\sim q_w(\phi)} ,
\end{align}
i.e., as an expectation value over flow-generated samples.

As pointed out in Ref.~\cite{Pawlowski:2022rdn}, the same idea can be applied to compute the DoS $\rho(c)$ defined in Eq.~\eqref{eqn:DoS1} and the gDoS $\rho(\beta,c)$ defined in Eq.~\eqref{eqn:gDoS1}. In these definitions, however, the integrands contain Dirac $\delta$-distributions that cannot be sampled directly. To make the expressions amenable to sampling, the $\delta$-distribution is therefore replaced by a Gaussian of finite width,
\begin{equation}\label{eqn:regdelta}
\delta(c-x)=\lim_{P\to\infty}\frac{1}{\mathcal{N}}e^{-\frac{P}{2}(c-x)^2}, 
    \qquad \mathcal{N}=\sqrt{\frac{2\pi}{P}},
\end{equation}
with the original constraint recovered in the limit $P\to\infty$.

For finite $P$, this replacement modifies the target action $S(\phi)$ in Eq.~\eqref{eqn:normDiffpartition}. One obtains an effective action $S_{c,P}(\phi)$ that depends on both the constraint $c$ and the regulator parameter $P$. The explicit form of the target actions relevant for this work will be given below in Eq.~\eqref{eqn:intensiveWilson} for the DoS and in Eq.~\eqref{eqn:targetTheta} for the gDoS.

Following Ref.~\cite{Pawlowski:2022rdn}, the resulting estimator for the regularized density of states is
\begin{equation}\label{eqn:flowDoS}
    \rho_P(c)
    = \int \mathcal{D}\phi\, e^{-S_{c,P}(\phi)}
    =\left\langle e^{-S_{c,P}(\phi)-\log q_{w,c}(\phi)} \right\rangle_{\phi\sim q_{w,c}(\phi)} ,
\end{equation}
where the NF generating the distribution $q_{w,c}(\phi)$ is trained separately for each value of the constraint parameter $c$. In this way, the density of states can be estimated directly from samples generated by the flow.

As discussed in Ref.~\cite{Pawlowski:2022rdn}, the estimator in Eq.~\eqref{eqn:flowDoS} offers a potential advantage over conventional MC-based approaches. Standard DoS methods typically determine only the derivative of the logarithm of the DoS, from which $\rho(c)$ must subsequently be reconstructed. In contrast, once an NF has been properly trained, it allows for a direct evaluation of $\rho_P(c)$ across the full range of the constrained variable. This can lead to highly precise determinations of the DoS over the entire domain, potentially providing improved control over relative errors, which is essential for reliably computing observables at arbitrary values of the coupling.

\section{U(1) lattice gauge theory in (1+1)D}\label{sec:3}

The (1+1)D U(1) lattice gauge theory provides a simple testbed for studying topology in lattice gauge theories. The theory contains only constant electric fields and no propagating degrees of freedom. Nevertheless, it has a well-defined topological charge, leading to distinct topological sectors. Additionally, introducing a $\theta$-term renders the action complex. As a result, Monte Carlo simulations suffer from critical slowing down near the continuum limit and from a sign problem induced by the $\theta$-term. The model therefore exhibits algorithmic challenges that also arise in lattice QCD, making it a useful benchmark for new numerical methods. In addition, in (1+1)D gauge theories are exactly solvable in the continuum~\cite{Migdal:1975zg}, and their lattice partition function is known analytically~\cite{Rusakov:1990rs}, providing a ground truth for comparison.

In the absence of a $\theta$-term, the U(1) lattice gauge theory is defined by the Wilson action (here written without the coupling $\beta$, which will be introduced later in Eqs.~(\ref{eqn:thetatermI}) and (\ref{eqn:intensiveDoSn})), 
\begin{equation}\label{eqn:WilsonAct}
    S_G[U] =  \sum_{n} \left[ \mathbb{I} - \frac{1}{2} \left(U_{\mu\nu}(n) + U^{\dagger}_{\mu\nu}(n) \right) \right],
\end{equation}
where the plaquette $U_{\mu\nu}(n) = U_{\mu}(n)\,U_{\nu}(n+\hat\mu)\,U^{\dagger}_{\mu}(n+\hat\nu)\,U^{\dagger}_{\nu}(n) $
is a gauge-invariant product of four link variables along a closed loop.

For a lattice of size $L\times L$ with periodic boundary conditions, the partition function is known exactly~\cite{Rusakov:1990rs}:
\begin{equation}\label{eqn:exactPF}
    Z(\beta) = e^{-\beta N_{\mathbb{P}}} \sum_{n \in \mathbb{Z}} \left[I_n (\beta) \right]^{N_{\mathbb{P}}},
\end{equation}
where $N_\mathbb{P}$ is the number of plaquettes and $I_n$ is the modified Bessel function of the first kind~\cite{Olver2010}.

Including a $\theta$-term generalizes the lattice action to
\begin{equation}\label{eqn:thetatermI}
S=\beta S_G(U)+iS_\theta(U)\,,\qquad S_{\theta}(U) = \theta Q_{\rm top}=\frac{\theta}{2 \pi}  \sum_{n} \text{arg}\left[ U_{\mu\nu}(n)\right],
\end{equation}
where $Q_{\rm top}$ is the geometric definition of the topological charge~\cite{Kanwar:2020xzo,Albergo:2021vyo}. This formulation has been studied previously with MC–based DoS methods~\cite{Gattringer:2015eey,Gattringer:2020mbf}, though with a different definition of $Q_{\rm top}$.

\section{Results} \label{sec:4}

We present preliminary results for two scenarios in (1+1)D U(1) lattice gauge theory: without a $\theta$-term, to benchmark the NF-based DoS approach against the exact partition function of Eq.~\eqref{eqn:exactPF}, and with a $\theta$-term, where the action becomes complex and a sign problem arises.

Throughout this work, we employ the NF ansatz of Ref.~\cite{Kanwar:2020xzo}, consisting of coupling-based  gauge-equivariant convolutional networks trained by minimizing the reverse Kullback-Leibler divergence (for details on the architecture, we refer the reader to Ref.~\cite{Kanwar:2020xzo}). This choice of NF architecture is motivated by the fact that the constrained distributions in the DoS formulation are derived from a target action that preserves gauge invariance, so a gauge-equivariant architecture ensures that this symmetry is respected by the learned model. 

In the following analysis, we adopt the notation of Ref.~\cite{Kanwar:2020xzo}, in which the flow is defined in terms of coupling blocks, each parameterized by a convolutional neural network with a fixed number of coupling layers, acting on local gauge-invariant quantities. For the $L=8$ lattice, our choice for the flow comprises 24 coupling blocks composed out of 2 coupling layers of width 16 and kernel size 7. For the smaller $L=4$ lattice, a reduced architecture with 20 coupling blocks made of 2 coupling layers of width 8 and kernel size 5 provides comparable accuracy.

\subsection{Results for (1+1)D U(1) gauge theory without a $\theta$-term}
For the U(1) gauge theory without a $\theta$-term, the DoS can be defined using an intensive version of the Wilson action~\cite{Bazavov:2012ex}, obtained by normalizing the gauge action by the number of plaquettes $N_{\mathbb{P}}$, which in (1+1)D is $N_{\mathbb{P}} = L^2$. The corresponding $P$-dependent target action reads
\begin{equation}
\label{eqn:intensiveWilson}
S_{c,P}(U) =  \frac{P}{2}\left[c-\frac{S_G(U)}{N_{\mathbb P}}\right]^2+\log{\mathcal{N}},
\end{equation}
where $\mathcal{N}$ is defined in Eq.~\eqref{eqn:regdelta}.
The DoS is evaluated using the NF-based estimator of Eq.~\eqref{eqn:flowDoS}:
\begin{equation}\label{eqn:intensiveDoSn}
\rho_{P}(c)
=
\int \mathcal{D}U \,
e^{-S_{c,P}(U)} =\left\langle e^{-S_{c,P}(U)-\log q_{w,c}(U)} \right\rangle_{U\sim q_{w,c}(U)} ,
\end{equation}
where the NF generating the distribution $q_{w,c}(U)$ is trained separately for each value of $c$. In the following, we analyze the dependence of the $P$-dependent DoS on $P$, and subsequently compare the reconstructed DoS to the exact result obtained via the inverse Laplace transform of Eq.~\eqref{eqn:exactPF}.

\begin{figure}[t]
    \centering
    \includegraphics[width=0.49\textwidth]{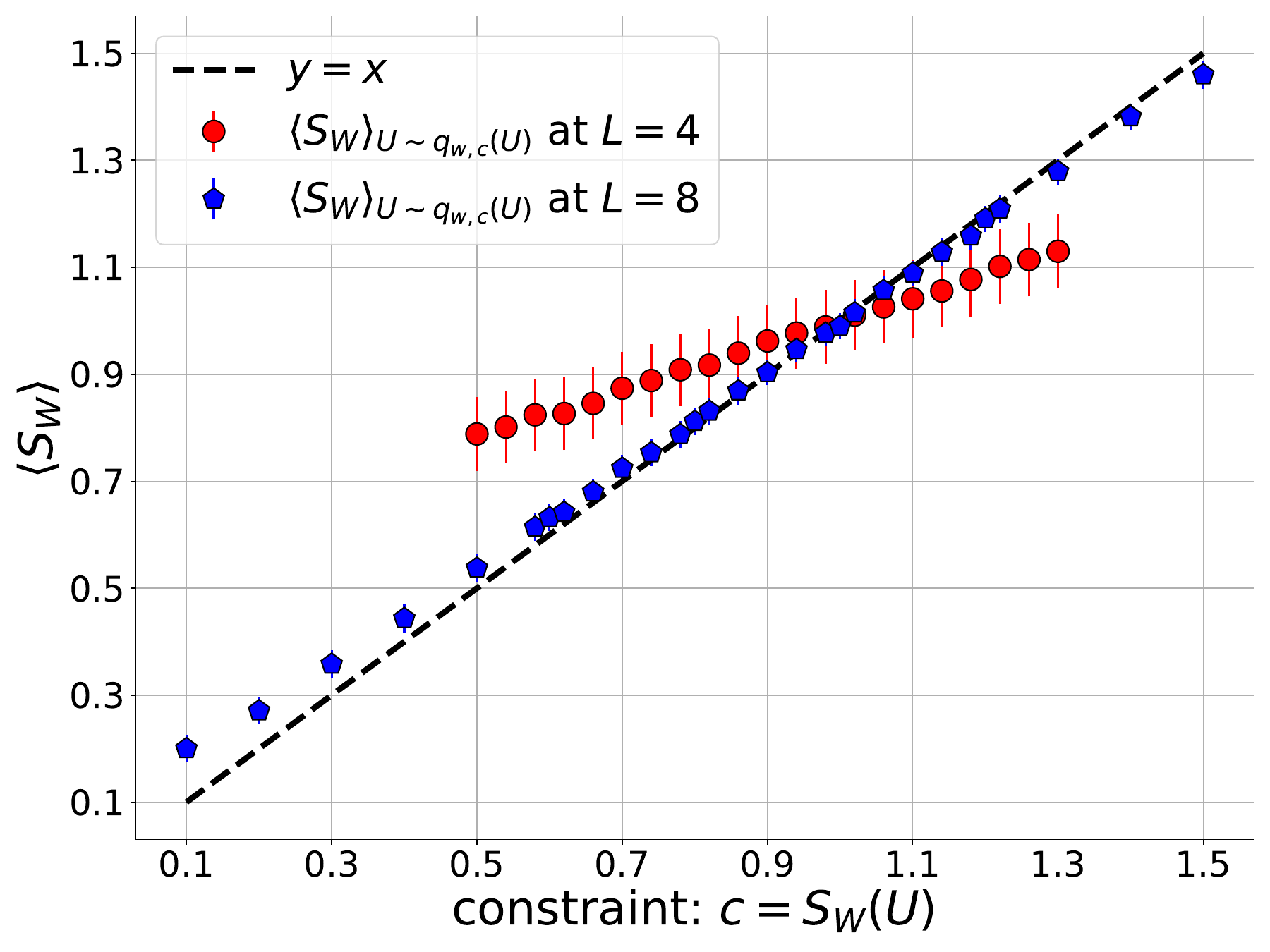}
    \hfill
     \includegraphics[width=0.49\textwidth]{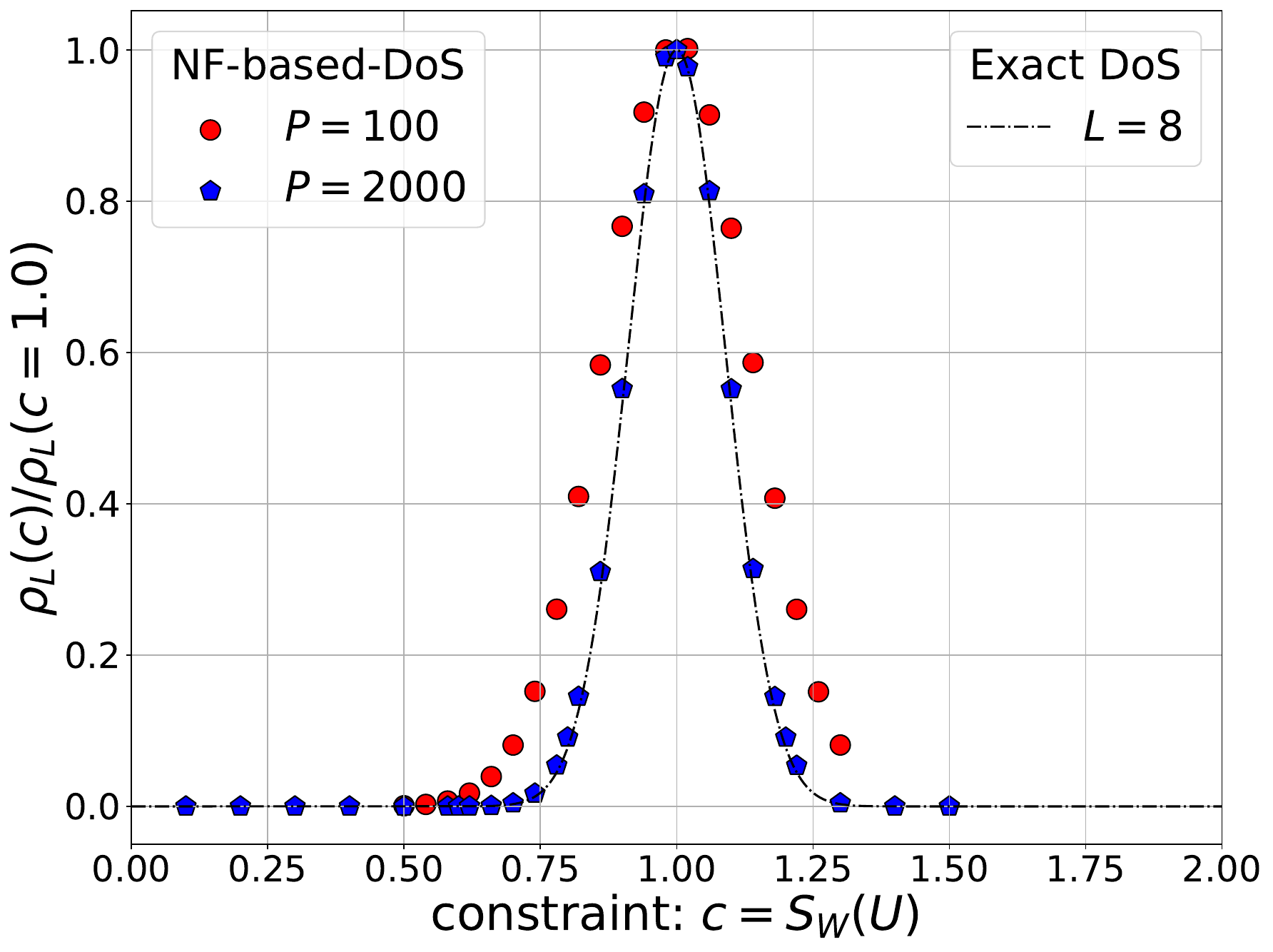}
    \caption{Results for (1+1)D U(1) gauge theory without a $\theta$-term obtained with the DoS method. Left: Expectation value of the Wilson action from NF-generated configurations at fixed constraint $c$ for an $L=8$ lattice and $P \in\{ 100, 2000\}$. The dashed line indicates perfect constraint satisfaction. Right: Reconstructed DoS compared to the exact result, illustrating that tighter constraints ($P=2000$) yield improved agreement.}
\label{fig:WilsonActionConstraintvsP}
\end{figure}

As a first validation of the constrained sampling procedure, we examine the relation between the imposed constraint $c$ and the expectation value of the Wilson action on the generated configurations. Figure~\ref{fig:WilsonActionConstraintvsP} (left) shows $\langle S_W \rangle_{U\sim q_{w,c}(U)}$ as a function of $c$ for an $L=8$ lattice at two values of the constraint strength, $P = 100$ and $P = 2000$. The dashed line $y=x$ indicates perfect constraint satisfaction, which would be obtained for a perfectly trained NF. We observe that the constraint is reproduced most accurately near $c \approx 1$, while deviations increase toward the boundaries of the constraint interval. This behavior is a direct consequence of the sampling measure defined in Eq.~\eqref{eqn:intensiveWilson}: configurations with extreme action densities are suppressed, and resolving the DoS in these regions requires enforcing a tighter constraint, i.e., increasing $P$.

The impact of constraint fidelity on the reconstructed DoS is shown in Fig.~\ref{fig:WilsonActionConstraintvsP} (right). For $P=100$, the NF-based estimates deviate substantially from the analytical result, whereas for $P=2000$, agreement is achieved across a wider range. We note that, at this stage, we do not provide uncertainty estimates for the DoS results. For the action results in Fig.~\ref{fig:WilsonActionConstraintvsP} (left) we included statistical sampling errors, in order to demonstrate the increase of statistical fluctuations with increasing width of the Gaussian introduced in Eq.~\eqref{eqn:regdelta}. However, these error bars do not include systematic effects, such as those coming from multiple NF training seeds. A detailed assessment of such systematic effects, both on the action results and the DoS results, will be addressed in future work.

\begin{figure}[t]
    \centering
    \includegraphics[width=0.49\textwidth]{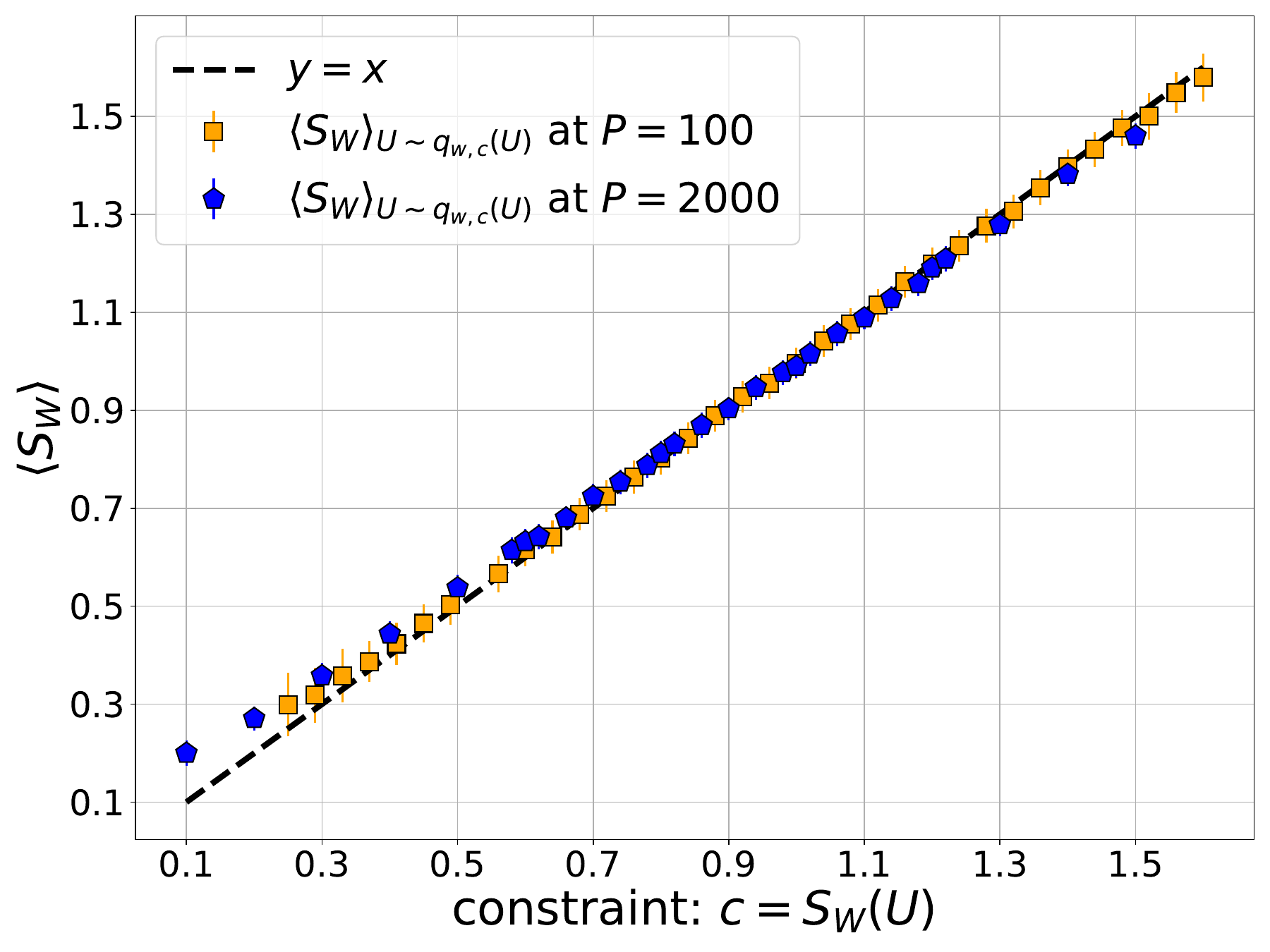}
    \hfill
     \includegraphics[width=0.49\textwidth]{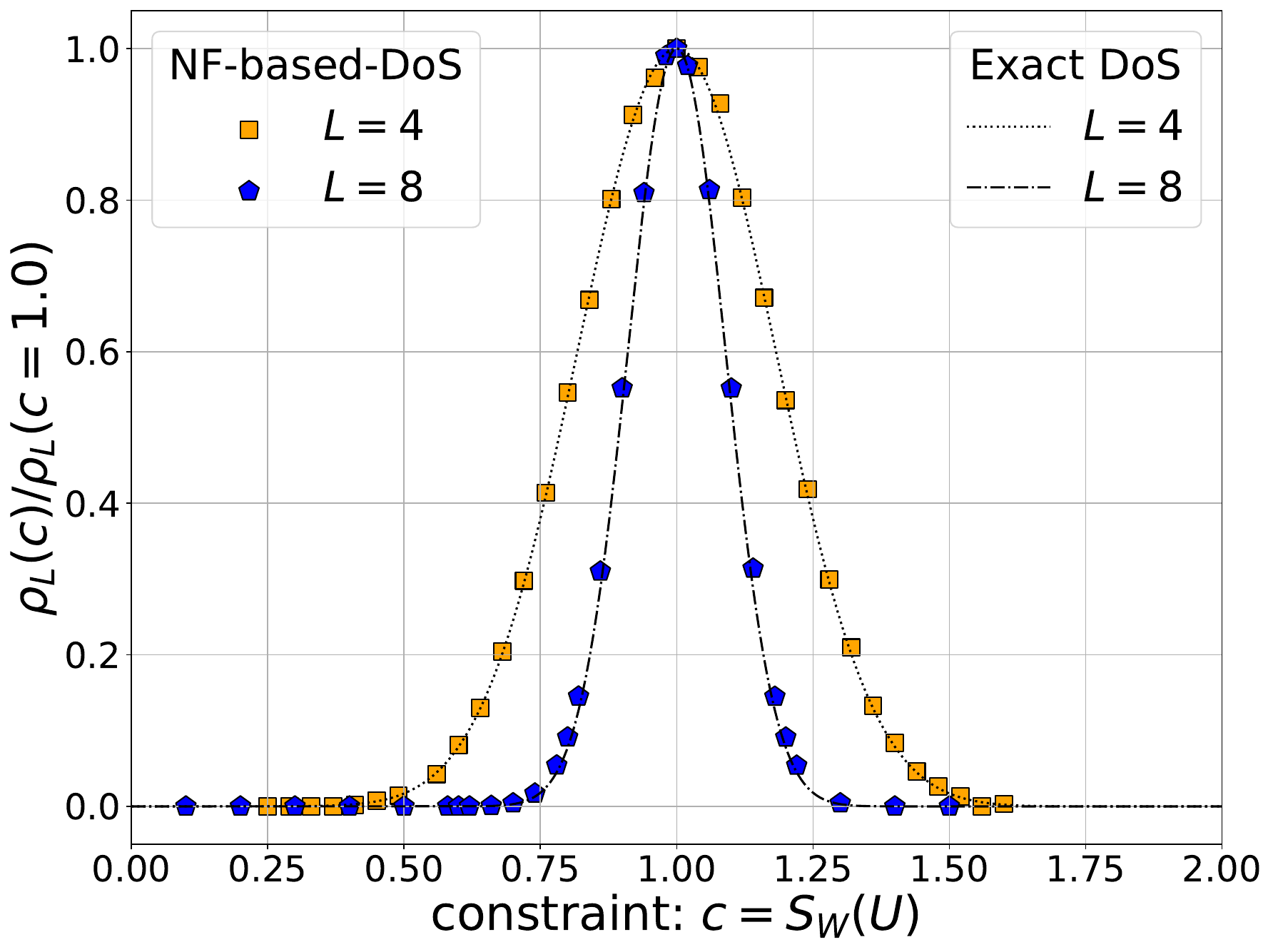}
    \caption{Results for (1+1)D U(1) gauge theory without a $\theta$-term obtained with the DoS method. Left: Expectation value of the Wilson action from NF-generated configurations at fixed constraint $c$ for $L = 4$ and $L = 8$ lattices with $P=2000$. The dashed line indicates perfect constraint satisfaction. Right: Reconstructed DoS compared to the exact result for both lattice sizes.}
    \label{fig:DoSU1}
\end{figure}

Motivated by these results, we fix the constraint strength to $P=2000$ and study two lattice volumes, $L \in \{4,8\}$. As illustrated in Fig.~\ref{fig:DoSU1}, the flow-based reconstruction qualitatively reproduces the exact DoS for both lattice sizes. Quantitatively, however, the precision is not yet sufficient to reliably compute derived observables, such as the topological susceptibility. In particular, for larger values of the gauge coupling $\beta$, the one-dimensional integrals defining the partition function become increasingly sensitive to small deviations in the DoS, since the Boltzmann factor amplifies contributions from specific regions of the action interval. This is reflected in the left panel of Fig.~\ref{fig:DoSU1}, where, despite the strong delta-like constraint, the measured action deviates from the imposed constraint near the edges of the interval.

To assess the quality of the learned distribution, we monitor the effective sample size (ESS), normalized to lie in $[0,1]$, with larger values indicating better overlap between the model and target distributions. We find that the ESS is highest in the central region of the action interval (reaching up to 72\% for $P=2000$, $L=8$, and $c=1.02$), where most configurations contribute, and up to an order of magnitude lower (down to 3.5\% for the same setup with $c=0.10$) near the interval boundaries. This indicates that the flow accurately captures the dominant regions of configuration space, while its performance deteriorates in regions where configurations are rare.

\subsection{Results for (1+1)D U(1) gauge theory with a $\theta$-term}

Next, we extend the analysis to the (1+1)D U(1) gauge theory in the presence of a $\theta$-term, testing the NF-based gDoS framework with the lattice definition of the topological charge given in Eq.~\eqref{eqn:thetatermI}. In this setting, the Euclidean action is $S(U) = \beta S_G(U) + i \theta Q_{\rm top}(U)$,
and the corresponding $P$-dependent target action for constrained sampling reads
\begin{equation}\label{eqn:targetTheta}
S_{c,P}(U) =  \beta S_G(U) + \frac{P}{2}\left[c-Q_{\mathrm{top}}(U)\right]^2+\log{\mathcal{N}},
\end{equation}
from which the DoS is evaluated using the NF estimator in Eq.~\eqref{eqn:intensiveDoSn}.

\begin{figure}[t]
    \centering
    \includegraphics[width=0.5\linewidth]{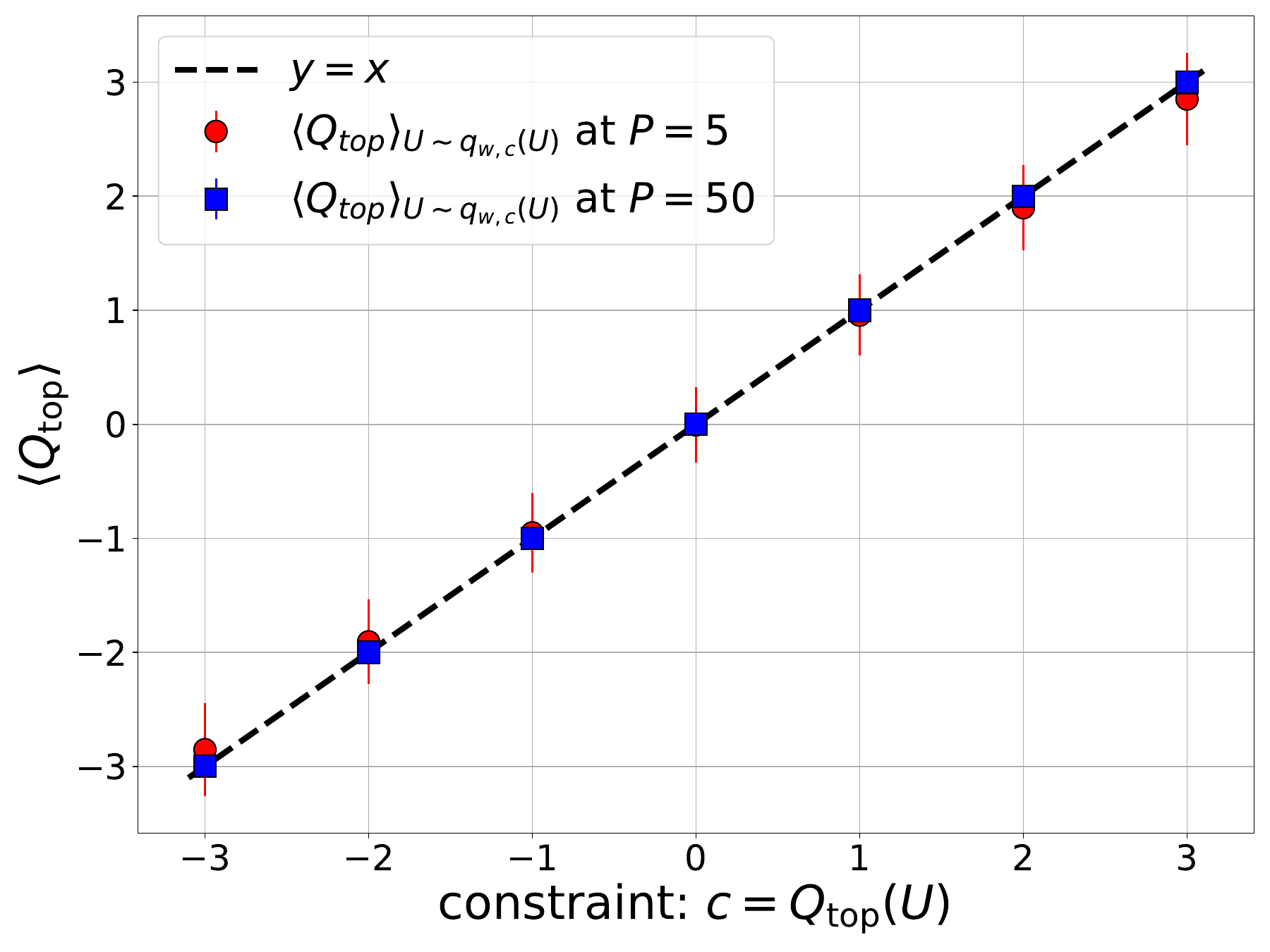}
    \caption{Results for (1+1)D U(1) gauge theory with a $\theta$-term obtained with the gDoS method. Verification of the topological charge constraint for an $L=8$ lattice. Shown is the expectation value of the topological charge from NF-generated gauge field configurations at fixed constraint, with $\beta=1.0$ and $P = \{5,50\}$. The dashed line indicates perfect constraint satisfaction.}
    \label{fig:topchargeconstraint}
\end{figure}

As in the pure Wilson case, we first verify that the imposed constraint effectively selects configurations with the desired topological charge. Using the geometric definition of $Q_{\rm top}$, the NF model is able to generate configurations corresponding even to relatively rare values of $Q_{\rm top}$. Figure~\ref{fig:topchargeconstraint} demonstrates that the constraint is accurately satisfied for moderate values of the constraint strength, $P = \{5,50\}$. This behavior reflects the fact that the DoS as a function of the topological charge is supported only at integer values, formally corresponding to a sum of Dirac $\delta$-functions~\cite{Gattringer:2020mbf}.

The primary objective of this analysis is to compute the gDoS at fixed $\beta$ and reconstruct the partition function as a function of $\theta$. Currently, the precision of this reconstruction is limited by the expressivity of the NF ansatz. In particular, we observe a pronounced reduction in the ESS when increasing $\beta$ or when constraining the system to configurations with large integer values of $Q_{\rm top}$. Work is ongoing to enhance the expressivity of the NF architecture to improve sampling efficiency and reconstruction accuracy.

\section{Conclusions and outlook}\label{sec:5}
In this preliminary work, we applied the NF-based DoS method to (1+1)D U(1) lattice gauge theory, both in the absence and in the presence of a $\theta$-term. We employed the gauge-equivariant NF ansatz of Ref.~\cite{Kanwar:2020xzo} and demonstrated that it can be used to sample from the $P$-dependent constrained action relevant for the DoS construction. In particular, in the presence of a $\theta$-term, this approach allows the generation of configurations at fixed values of the topological charge.

We also investigated the effect of relaxing the Dirac-delta constraint appearing in the DoS definition. Our results indicate that accurately learning the constrained distributions in regions where configurations are rare requires a sufficiently tight constraint. At the same time, increasing the constraint strength $P$ reduces the ESS of the flow model, reflecting a decreased overlap between the learned and target distributions in these regions.

This behavior suggests that the current limitation arises primarily from the expressivity of the NF architecture, particularly in the tails of the constrained distributions where configurations contribute with small probability but remain important for an accurate reconstruction of the DoS. Improving the expressivity of the flow model therefore appears to be a promising direction for achieving the precision required to reliably extract physical observables.

More generally, our results indicate that NF architectures designed to sample from the Boltzmann distribution of a given lattice field theory can be naturally adapted to the DoS framework, since the symmetry structure of the action is preserved under the imposed constraints. We are currently exploring this strategy in other systems, including the Hubbard model \cite{Freche_work_in_progress}, where more expressive flow architectures have been shown to improve sampling efficiency~\cite{Schuh:2025fky,Schuh:2025gks,Kreit:2026eng}.

\acknowledgments
\noindent
The authors thank Julian Urban and Aishwarya Deshpande for discussions during the initial phase of this project, and Felicitas Freche, Dominic Schuh, Janik Kreit, Emil Rosanowski, Andrea Bulgarelli, and Nico Dichter for ongoing discussions. The authors are especially grateful to Dominic Schuh for valuable inputs on these proceedings. We gratefully acknowledge the access to the Marvin cluster of the University of Bonn. This project was supported by the Deutsche Forschungsgemeinschaft (DFG, German Research Foundation) as part of the CRC 1639 NuMeriQS -- project no.\ 511713970. The code used for the results in this paper is adapted from Ref.~\cite{Albergo:2021vyo}.

\bibliographystyle{JHEP}
\bibliography{refs}

\end{document}